\documentclass[sigconf]{acmart}

\usepackage[utf8]{inputenc}
\usepackage{colortbl}
\hypersetup{hidelinks}
\usepackage{pifont}%
\usepackage{balance}
\usepackage{dirtytalk}
\usepackage{enumitem}

\newcommand{\cn}{\textsc{MultiMediate\,}}

\AtBeginDocument{%
  \providecommand\BibTeX{{
    \normalfont B\kern-0.5em{\scshape i\kern-0.25em b}\kern-0.8em\TeX}}}

\copyrightyear{2026}
\acmYear{2026}
\setcopyright{cc}
\setcctype{by}
\acmConference[MM '26]{Proceedings of the 34th ACM International Conference on Multimedia}{November 10--14, 2026}{Rio de Janeiro, Brazil.}
\acmBooktitle{Proceedings of the 34th ACM International Conference on Multimedia (MM '26), November 10--14, 2026, Rio de Janeiro, Brazil}
\acmDOI{10.1145/3767308.3837718}
\acmISBN{979-8-4007-2213-4/2026/11}

\begin{document}

\author{Marius Funk}
\affiliation{
    \institution{University of Augsburg}
    \city{Augsburg}
    \country{Germany}
}
\affiliation{
    \institution{JAIST}
    \city{Nomi, Ishikawa}
    \country{Japan}
}
\email{marius.funk@uni-a.de}

\author{Daksitha Withanage Don}
\affiliation{
    \institution{University of Augsburg}
    \city{Augsburg}
    \country{Germany}
}
\email{daksitha.withanage.don@uni-a.de}

\author{Michal Balazia}
\affiliation{
    \institution{INRIA Université Côte d'Azur}
    \city{Sophia Antipolis}
    \country{France}
}
\email{michal.balazia@inria.fr}

\author{Victor Oei}
\affiliation{
    \institution{University of Stuttgart}
    \city{Stuttgart}
    \country{Germany}
}
\email{victor.oei@vis.uni-stuttgart.de}

\author{Shogo Okada}
\affiliation{
    \institution{JAIST}
    \city{Nomi, Ishikawa}
    \country{Japan}
}
\email{okada-s@jaist.ac.jp}

\author{Fran\c{c}ois Br\'emond}
\affiliation{
    \institution{INRIA Université Côte d'Azur}
    \city{Sophia Antipolis}
    \country{France}
}
\email{francois.bremond@inria.fr}

\author{Andreas Bulling}
\affiliation{
    \institution{University of Stuttgart}
    \city{Stuttgart}
    \country{Germany}
}
\email{andreas.bulling@vis.uni-stuttgart.de}

\author{Elisabeth Andr\'e}
\affiliation{
    \institution{University of Augsburg}
    \city{Augsburg}
    \country{Germany}
}
\email{elisabeth.andre@uni-a.de}

\author{Philipp M\"uller}
\affiliation{
    \institution{MPI for Intelligent Systems}
    \city{Stuttgart}
    \country{Germany}
}
\email{mueller@is.mpg.de}

\title[The \cn'26 Challenge]{Engagement Estimation in Human-Robot and Human-Human Interactions across Domain, Age and Culture:\\ The \cn'26 Challenge}

\renewcommand{\shortauthors}{Funk et al.}

\begin{abstract}

Understanding engagement in human interaction is essential for modelling assistive and context-aware AI systems. Previously,  the \cn challenge has focused on estimating momentary 
continuous 
engagement values in adult conversation across cultures, achieving high levels of generalisability across domains. However, engagement can be measured in different metrics and manifest differently between children and adults.
\cn'26 therefore expands the evaluated domain set to include the \textsc{PInSoRo} dataset of Child-Child and Child-Robot play sessions in English. These sessions were annotated categorically with task and social engagement, diversifying engagement types, age ranges and interaction types. 
In this paper, we describe the \cn'26 challenge and its goals, introduce the \textsc{NoXi}, \textsc{NoXi-J}, \textsc{MPIIGI}, and \textsc{PInSoRo} datasets, showcase extracted multi-modal features, present our baseline model for engagement estimation across all domains, and discuss challenge submissions and approaches. In an effort to provide a transparent and welcoming challenge environment,
baseline code, leaderboard, datasets, and extracted features are publicly available at \url{https://github.com/hcmlab/MultiMediate_26}.

\end{abstract}



\maketitle

\section{Introduction}

Estimating the momentary degree of user engagement is crucial for AI systems that take part in and support social interaction.
Consequently, a growing body of work proposed approaches for the automatic estimation of engagement in human-human and human-machine interactions.

Engagement in human-human interactions was predicted using gaze features \cite{Bednarik:2012:GCE:2401836.2401846Engagement,NakanoEstimating} and multi-modal fusion approaches incorporating audio signals, gaze, eye contact, facial features and dialogue transcriptions \cite{ma2025multimodal,li2023data,yang2023multimediate,li2024dat, yu2025multimediate, yu2025generalizable}.
Human-agent and human-robot interactions were also analysed using backchannels \cite{CharlesRichEngagementHumanRobotInteraction}, body posture \cite{Sanghvi:2011:AAA:1957656.1957781BodyMotion} and multi-modal 
transformer models \cite{Lim2025}. 
Research also showed success in predicting engagement in children \cite{rajagopalan_play_2015}, including interactions with artificial agents \cite{park_model-free_2019, jain_modeling_2020}.

\begin{figure*}[t]
\centering
\includegraphics[width=0.97\textwidth]{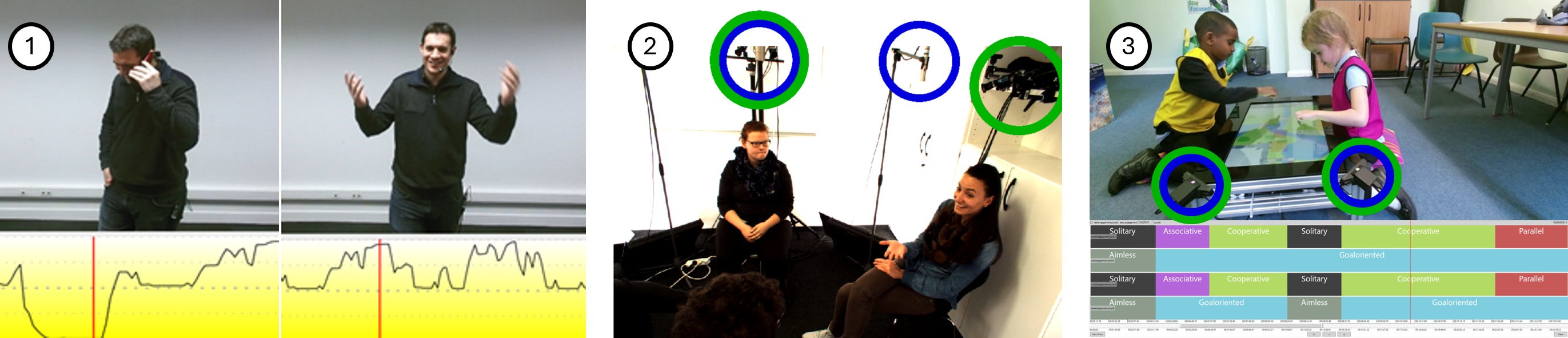}
\caption{(1) Examples of low (left) and high (right) engagement in the NoXi corpus \cite{cafaro:et:al:2017}. (2) Setup of the MPIIGroupInteraction dataset~\cite{muller_detecting_2018}. (3) Annotated Child-Child free-play session of the PInSoRo dataset \cite{lemaignan2018pinsoro} displayed using the NOVA tool \cite{Baur2020}. All images were used with the permission of the respective authors.}
\Description{Image is split into three parts: (1) Examples of low (left) and high engagement (right) in the NoXi corpus \cite{cafaro:et:al:2017}. (2) Setup of the MPIIGroupInteraction dataset~\cite{muller_detecting_2018}. (3) Annotated Child-Child free-play session of the PInSoRo dataset \cite{lemaignan2018pinsoro} displayed using the NOVA tool \cite{Baur2020}.}
\label{fig:datasets}
\end{figure*}

\cn is an expanding yearly challenge that focuses on human behaviour analysis in social interaction. It has addressed eye contact detection \cite{muller2021multimediate}, backchannel detection \cite{muller2022multimediate}, and backchannel recognition \cite{muller2023multimediate}, while recent years have focused on the creation of engagement estimation approaches that are generalisable across cultures and conversational settings~\cite{muller2024multimediate,withanage2025multimediate}.
While these efforts have led to significant advances in engagement estimation, they remain constrained to a specific age group (adult), and a single notion of engagement (single overall magnitude per time point).

\begin{table}[!t]
\centering
\begingroup
\footnotesize
\setlength{\tabcolsep}{5pt}
\renewcommand{\arraystretch}{0.95}
\caption{\cn'26 datasets: Languages and session count by dataset and split. $R$ denotes continuous and $C$ categorical engagement labels.}
\label{tab:data_overview}
\begin{tabular}{@{}p{2.4cm}p{1.6cm}p{1.7cm}p{1.6cm}@{}}
\toprule
\textbf{Dataset} & \textbf{Training Data} & \textbf{Validation Data} & \textbf{Test Data} \\
\midrule
\textbf{NoXi} ($R$)~\cite{cafaro:et:al:2017} 
& \emph{English (23), \newline French (7), \newline German (8)}
& \emph{English (3), \newline French (4), \newline German (3)}
& \emph{English (6), \newline French (6), \newline German (4)} \\
\vspace{0.1cm} \\
\textbf{NoXi Add.} ($R$)~\cite{cafaro:et:al:2017} 
 & - & - & \emph{Indonesian (4),  \newline Spanish (4), \newline Italian (2), \newline Arabic (2)} \\
\vspace{0.1cm} \\
\textbf{NoXi-J} ($R$)~\cite{funk2024noxij} &
\emph{Japanese (21), \newline Chinese (10)}
& \emph{Japanese (6), \newline Chinese (4)}
& \emph{Japanese (6), \newline Chinese (4)} \\
\vspace{0.1cm} \\
\textbf{MPIIGI} ($R$)~\cite{muller_detecting_2018}
& -  & \emph{German (6)}
& \emph{German (6)} \\

\vspace{0.1cm} \\
      \textbf{PInSoRo CC} ($C$)~\cite{lemaignan2018pinsoro} 
      & {\textit{English (20)}} 
      & {\textit{English (7)}} 
      & {\textit{English (6)}} \\
\vspace{0.1cm} \\
      \textbf{PInSoRo CR} ($C$)~\cite{lemaignan2018pinsoro} 
     & {\textit{English (12)}} 
     & {\textit{English (5)}} 
     & {\textit{English (6)}} \vspace{0.1cm} \\
\bottomrule
\end{tabular}
\endgroup
\end{table}

\cn'26 significantly broadens the challenge scope by including the \textsc{PInSoRo} dataset \cite{lemaignan2016towards,lemaignan2018pinsoro}, a corpus of dyadic Child-Child and Child-Robot play sessions. 
These sessions differ from the established estimation task in significant ways. First, they show interactions of children with each other and with robots with only sparse adult participation. Second, they are not conceived as conversations, but as free play interactions with varying amounts of speech. Finally, engagement was not annotated using a continuous scale, but through two sets of categorical annotation labels: task engagement and social engagement. 

\section{Challenge Description}

In \cn'26, we focus on a multi-domain engagement estimation challenge across age groups, cultures, and engagement labels using a variety of datasets.  Participating teams submit their predictions for the test set to an EvalAI server\footnote{https://challenges.hcai.eu/web/challenges/challenge-page/29/overview} for up to 20 submissions per team. We also continue our established challenge tasks of Bodily Behaviour Recognition, Backchannel Detection and Eye Contact Detection using the same server submission infrastructure.

\subsection{Task Definition}

The task is to accurately predict different representations of engagement. We measure this using the Concordance Correlation Coefficient (CCC) ~\cite{ccc}, ranging from \(-1\) to \(+1\) for the datasets using continuous engagement annotation, namely \textsc{NoXi}, \textsc{NoXi-J}, \textsc{NoXi (Add.)}, and \textsc{MPIIGI}. \cn'26 additionally evaluates categorical social and task engagement for the \textsc{PInSoRo} dataset using Cohen's Kappa ($\kappa$), which also ranges from \(-1\) to \(+1\) and is therefore comparable to CCC. Using the average CCC score ($\mathrm{CCC}_{\mathrm{Avg}}$) and the average $\kappa$ score ($\mathrm{\kappa}_{\mathrm{Avg}}$), the overall score is calculated as:

\begin{equation}
    \mathbf{Global}= \frac{2}{3}\,\mathrm{CCC}_{\mathrm{Avg}} + \frac{1}{3}\,\mathrm{\kappa}_{\mathrm{Avg}}
    \label{eq:score_cal}
\end{equation}

\subsection{Datasets}

\cn'26 encompasses a wide variety of datasets and frame-wise engagement labels. The challenge contains recordings of interactions in nine languages for dyadic conversations, group interactions, and child play sessions including Child-Child and Child-Robot interactions. We continue the use of the \textit{NO}vice e\textit{X}pert \textit{I}nteraction database (\textsc{NoXi})~\cite{cafaro:et:al:2017} with its additional languages, \textsc{NoXi (Add.)}, \textsc{NoXi-J}, the extension of \textsc{NoXi} recorded in \textit{J}apan \cite{funk2024noxij}, and the MPIIGroupInteraction (\textsc{MPIIGI}) dataset~\cite{muller_detecting_2018}.  An overview can be seen in Table~\ref{tab:data_overview}. The datasets and all extracted features are available on the challenge website\footnote{\url{https://multimediate-challenge.org/Dataset/}}. For the PInSoRo dataset, we share only extracted features.

This year, we extend our challenge with the Plymouth Interacting Social Robots (\textsc{PInSoRo})
dataset containing recordings of English-language free-play sessions in the form of Child-Child (20/7/6) and Child-Robot (12/5/6) interactions with an average duration of \(\sim\)39 minutes \cite{lemaignan2016towards,lemaignan2018pinsoro}. The unstructured interactions (see Figure \ref{fig:datasets}, part 3) are dyadic with adult supervision, co-located around a touchscreen table, and were annotated categorically for social and task engagement \cite{lemaignan2018pinsoro}:
\begin{enumerate}
    \item \textit{Task engagement.} The child's engagement in the free-play task divided into four categories: \textit{goal-oriented}, \textit{aimless}, \textit{adult-seeking}, and \textit{no play}.
    \item \textit{Social engagement.} The child's social engagement divided into categories based on five of Parten's stages of play \cite{parten1932social}: \textit{solitary}, \textit{onlooker}, \textit{parallel}, \textit{associative}, and \textit{cooperative}.
\end{enumerate}
All sessions of the validation and test sets are annotated by at least two raters. 
Besides annotations provided by \citet{lemaignan2018pinsoro}, ten additional CR sessions were annotated to guarantee this paradigm.
We provide all available annotations of the train/validation set sessions. The test set is evaluated exclusively on the frames for each subtask in which all annotators agree on the label.

\paragraph{Training data policy.}

Participants can use the provided data in any configuration for model training (single multi-headed model, multiple models, domain adaptation). We ask participants to share their chosen approach for transparency. They are  allowed to incorporate additional external open-source corpora during training.

\section{Feature Extraction}

Video, audio and text features were extracted from all challenge datasets and provided to the challenge participants. We do not share \textsc{PInSoRo} videos or transcripts due to privacy constraints.

\subsection{Visual Features}

\begin{itemize}[leftmargin=0pt, itemindent=1em]
    \item[] \textit{Handcrafted Features}:
We used \textit{OpenFace~2.0} \cite{baltrusaitis2018openface} (714 Dim.) to extract gaze, facial landmarks and features, head pose, and action units as well as \textit{OpenFace~3.0} \cite{hu2025openface} (21 Dim.) to extract eight action units, gaze direction and emotional labelling. \textit{OpenPose} was used for 2-D pose estimation of hand and body keypoints \cite{cao2017realtime} (139 Dim.). All features were extracted frame-wise.

    \item[] \textit{Visual Embeddings}:
\textit{DINOv2}'s ViT-B model \cite{oquab2024dinov} (768 Dim.), \textit{Swin Transformer}  \cite{liu2021Swin,VideoSwinTransformer}  (768 Dim.), and \textit{VideoMAE~v2} \cite{videomaev2}  (1408 Dim.) were used to extract visual embeddings with a stride of 16 frames. A frame-wise embedding space of joint text and image was derived via \textit{CLIP} \cite{radford2021learning} (512 Dim.).

\end{itemize}

\subsection{Audio and Text Features}

Audio features include handcrafted and embedding-space auditive features as well as transcriptions and sentence embeddings. All features were extracted using the DISCOVER pipeline \cite{schiller2024discover} with a 1\,s window size and a 40\,ms stride (except for the transcript). \textbf{NoXi} and \textbf{NoXi-J} audio is diarised by default as per the recording setup. Features for  \textbf{MPIIGI} were extracted after human-annotated speaker diarisation. No diarisation was performed for \textsc{PInSoRo} audio features.

\begin{itemize}[leftmargin=0pt, itemindent=1em]
    \item[] \textit{Audio features}: We extracted a variety of frequency, energy and spectral parameters using the handcrafted extended Geneva Minimalistic Acoustic Parameter Set (\textit{eGeMAPS v2}) \cite{eyben2015geneva, valstar2016avec} (88 Dim.). To complement the handcrafted features, we also provide the \textit{w2v-BERT 2.0} model's embedding-space output \cite{barrault2023seamless} (1024 Dim.).

    \item[] \textit{Text features}: All speech was transcribed using \textit{WhisperX} \cite{bain2022whisperx,radford2023robust}. 
    Multilingual embeddings were then extracted from each transcribed utterance by \textit{XLM-RoBERTa} \cite{unsupervised2019conneau} (768 Dim.).
\end{itemize}

\section{Baseline Prediction}

\begin{table}[t]
\centering
\footnotesize
\caption{Test set baseline selection for \textit{continuous} engagement labels. 
}
\label{tab:continuous_baseline}
\setlength{\tabcolsep}{3pt}
\begin{tabular}{lccccc}
\toprule
\textbf{Feature Set} & \textsc{NoXi}  & \textsc{NoXi-J} & \textsc{MPIIGI} & \textsc{NoXi (Add.)} & \textbf{$\mathrm{CCC}_{\mathrm{Avg}}$} \\
\midrule
\multicolumn{6}{l}{\textit{Video}} \\
\quad OpenFace 2.0    & 0.11  & 0.16 & 0.12 & 0.19 & 0.14 \\
\quad OpenFace 3.0    & 0.22 & 0.27 & 0.09 & 0.11 & 0.17 \\
\quad OpenPose        & 0.48 & 0.27 & 0.09 & 0.43 & 0.32 \\
\quad CLIP            & 0.44 & 0.22 & -0.01 & 0.37 & 0.25 \\
\quad DINOv2            & 0.53 & 0.11 & 0.01 & 0.46 & 0.28 \\
\quad Swin Transformer & 0.54 & 0.26 & -0.05 & 0.46 & 0.31 \\
\quad VideoMAE        & 0.61 & 0.22 & 0.02 & 0.56 & 0.35 \\
\midrule
\multicolumn{6}{l}{\textit{Audio}} \\
\quad \textbf{eGeMAPS v2}      & \textbf{0.55} & \textbf{0.31} & \textbf{0.45} & \textbf{0.49} & \textbf{0.45} \\
\quad w2v-BERT 2.0        & 0.60 & 0.32 & 0.05 & 0.47 & 0.36 \\
\midrule
\multicolumn{6}{l}{\textit{Text}} \\
\quad XLM-RoBERTa     & 0.39 & 0.26 & 0.01 & 0.27 & 0.23 \\
\bottomrule
\end{tabular}
\end{table}

\begin{table}[t]
\centering
\footnotesize
\caption{Test set baseline selection for \textit{categorical} engagement labels. 
}
\label{tab:categorical_baseline}
\setlength{\tabcolsep}{3pt}
\begin{tabular}{lccccc}
\toprule
\textbf{Feature set} & \textbf{CC Social}& \textbf{CC Task}& \textbf{CR Social} & \textbf{CR Task}& \textbf{$\mathrm{\kappa}_{\mathrm{Avg}}$} \\
\midrule
\multicolumn{6}{l}{\textit{Video}} \\
\quad OpenFace 2.0    & 0.08 & 0.06 & -0.01 & 0.25 & 0.09 \\
\quad OpenFace 3.0    & 0.13 & 0.10 & 0.02  & 0.17 & 0.10 \\
\quad OpenPose        & 0.06 & 0.03 & -0.03 & 0.21 & 0.07 \\
\quad CLIP            & 0.09 & 0.12 & -0.02 & 0.16 & 0.09 \\
\quad DINOv2            & 0.08 & 0.07 & 0.01  & 0.08 & 0.06 \\
\quad \textbf{Swin Transformer} & 0.09 & 0.08 & \textbf{0.09} & \textbf{0.24} & \textbf{0.13} \\
\quad VideoMAE        & 0.14 & 0.13 & -0.02 & 0.05 & 0.08 \\
\midrule
\multicolumn{6}{l}{\textit{Audio}} \\
\quad eGeMAPS v2      & 0.13 & 0.02 & -0.02 & -0.03 & 0.03 \\
\quad w2v-BERT 2.0        & 0.11 & 0.05 & 0.00  & 0.03  & 0.05 \\
\midrule
\multicolumn{6}{l}{\textit{Text}} \\
\quad \textbf{XLM-RoBERTa}     & \textbf{0.14} & \textbf{0.16} & -0.03 & 0.10 & 0.10 \\
\bottomrule
\end{tabular}
\end{table}

We use a feed-forward MLP with three ReLU hidden layers and dropout after the second hidden layer based on the \cn'25 baseline architecture \cite{withanage2025multimediate}. 
The hyperparameters were determined using KerasTuner Hyperband~\cite{omalley2019kerastuner} with log uniform sampling over hidden widths, batch size, dropout rate and learning rate.
All models are trained using the Adam optimiser.

\begin{table*}[!t]
  \centering
  \caption{Results of the multi-domain engagement estimation task. 
  }
  \label{tab:engagement-leaderboard}
  \footnotesize
  \setlength{\tabcolsep}{2.5pt}
  \begin{tabular}{lccccccccccc}
    \toprule
    Team & NoXi & NoXi-J & MPIIGI & NoXi (Add.) & $\mathrm{CCC}_{\mathrm{Avg}}$ & CC Social & CC Task & CR Social & CR Task & $\mathrm{\kappa}_{\mathrm{Avg}}$ & Global \\
        \midrule
    \multicolumn{12}{l}{\textit{Participants MM 2026}} \\
    \quad \textbf{HFUT-LMC}        & 0.80 & 0.59 & 0.67 & 0.77 & 0.71 & \textbf{0.65} & \textbf{0.48} & \textbf{0.46} & 0.51 & \textbf{0.52} & \textbf{0.65} \\
    \quad NUDT\_LDM-AF    & \textbf{0.82} & 0.60 & 0.71 & \textbf{0.80} & 0.73 & 0.37 & 0.46 & 0.41 & \textbf{0.58} & 0.46 & 0.64 \\
    \quad mltac           & 0.81 & \textbf{0.66} & 0.60 & 0.73 & 0.70 & 0.33 & 0.35 & 0.25 & 0.51 & 0.36 & 0.59 \\
    \quad USTC\_IAT\_United & 0.81 & 0.64 & \textbf{0.75} & 0.76 & \textbf{0.74} & 0.23 & 0.20 & 0.23 & 0.24 & 0.22 & 0.57 \\
    \midrule
    \multicolumn{12}{l}{\textit{Winner MM 2025}} \\
    \quad HFUT-LMC        & 0.79 & 0.58 & 0.67 & 0.75 & 0.70 & -- & -- & -- & -- & -- & -- \\
    \midrule
    \midrule
    \multicolumn{12}{l}{\textit{Baseline (MM 2026)}} \\
    \quad Baseline (ours) & 0.55 & 0.31  & 0.45 & 0.49   & 0.45 & 0.14 & 0.16 & 0.09 & 0.24 & 0.16 & 0.36 \\
    \bottomrule
\end{tabular}
\end{table*}

\subsection{Regression Models} 
The models estimating continuous labels were trained using a regression output layer and MSE loss with RMSE used as the performance metric. Test performance is reported as CCC. Based on the \cn'25 baseline results, two models were trained on the NoXi and NoXi-J train and validation sets respectively. The model trained on \textsc{NoXi} was evaluated on the hold-out test sets for \textsc{NoXi}, \textsc{NoXi (Add.)}, and \textsc{MPIIGI}. The \textsc{NoXi-J} test set was evaluated using the model trained on \textsc{NoXi-J}.

\subsection{Categorical Models} 
Models estimating categorical labels use a multi-head approach for task and social engagement, the same trunk as the regression model, and a combined categorical cross-entropy loss. Per-sample weights were computed to account for class frequency. Two models were trained on the \textsc{PInSoRo} Child-Child (CC) and Child-Robot (CR) interactions train and validation sets and evaluated on their respective test set using Cohen's Kappa ($\kappa$) \cite{cohen1960coefficient}.

\section{Results}

\subsection{Baseline Models}

We trained models for each extracted feature on the  train and validation splits of \textsc{NoXi},  \textsc{NoXi-J},  \textsc{PInSoRo CC}, and  \textsc{PInSoRo CR}. Regression and cross-entropy model results are reported separately and then combined into the overall score. 
To include all datasets, we report the per-feature test scores. The chosen baseline should be seen as an upper bound on the introduced architecture.

\paragraph{Regression models} The overall best performing feature for the regression model was the \textit{eGeMAPS v2} audio feature with an average CCC of 0.45 (see Table \ref{tab:continuous_baseline}). It showed cross-domain generalisability with comparatively high performance for both in-domain (\textsc{NoXi}: 0.55, \textsc{NoXi-J}: 0.31) and out-of-domain (\textsc{MPIIGI}: 0.45, \textsc{NoXi Add.}: 0.49) test data. 

\paragraph{Categorical models}  The categorical models showed strong differences in performance between CC and CR interactions (see Table \ref{tab:categorical_baseline}). While \textit{XLM-RoBERTa} performed best for the CC subset with an average $\kappa$ of 0.15 (CC Social: 0.14, CC Task: 0.16), the model underperformed for CR interactions (CR Social: -0.03, CR Task: 0.10). This mediocre performance was expected: in contrast to the lively CC sessions, most CR interactions consist of the child playing silently with the robot.  \textit{Swin Transformer} showed the best performance on the CR dataset (CR Social: 0.09, CR Task: 0.24), as well as the best overall performance with a $\kappa$ of 0.13. The structural differences in interaction %
motivated us to choose the \textit{XLM-RoBERTa} trained model for the CC baseline and the \textit{Swin Transformer} model for the CR baseline.

\paragraph{Overall score} We combine the prediction results of \textit{eGeMAPS v2} for the regression models ($\mathrm{CCC}_{\mathrm{Avg}}$: 0.45) and the average of the \textit{XLM-RoBERTa} CC model performance and  \textit{Swin Transformer} CR model performance ($\mathrm{\kappa}_{\mathrm{Avg}}$: 0.16). We calculate the overall score with the formula defined in Eq. \ref{eq:score_cal} (Global: 0.36).

\subsection{Challenge Solution Results}
\label{sec:challenge-results}

We had 13 participating groups in the multi-domain engagement estimation challenge. The final leaderboard showing all published papers ranked by the global score can be seen in Table~\ref{tab:engagement-leaderboard}. 

HFUT-LMC~\cite{yu2026semantic} won the challenge with a Semantic-Enhanced Sparse Temporal framework (SEST) that models engagement-related information from semantic and temporal perspectives using multi-modal LLMs. They achieved a global score of 0.65 and the best average on the new PInSoRo dataset (0.52), performing best on all CC interaction tasks (Social: 0.65, Task: 0.48) and the CR social task (0.46). The model also outperforms the regression task score of DAPA (Domain-Adaptive Parallel Attention), their own domain prompting approach and last year's challenge winner \cite{yu2025generalizable}. 

The best performance on last year's tasks, however, was achieved by USTC\_IAT\_United~\cite{zou2026uniee} (0.74), who extended the DAPA model by structured multi-modal scaling and multi-party interaction generalisation, showing excellent performance in the few-shot adaptation to the \textsc{MPIIGI} dataset (0.75). 

The most accurate NoXi (0.82) and NoXi (Add.) (0.80) predictions were achieved by NUDT\_LDM-AF \cite{deng2026ldmaf}, who used Learnable Domain Memories and Adaptive Filtering (LDM-AF) to enable the model to efficiently learn from multiple domains. The best NoXi-J performance (0.66) was achieved by mltac \cite{merrild2026engagement}  using Temporal Convolutional Networks and interaction modelling.

While these submissions outperformed the winner of last year's regression task, the progress seems to be plateauing, with NoXi only improving by 0.03 points. 
The newly introduced categorical prediction subtask exhibited considerable variation in model performance, indicating substantial potential for further advancement in future research.

The previous challenge tasks showed minor improvements: \textit{Bodily Behaviour Recognition} accuracy improved from 0.65 to 0.67, \textit{Eye Contact Detection} from 0.82 to 0.83. \textit{Backchannel Detection} received one submission that did not improve accuracy.

\section{Conclusion}

\cn'26 expanded the engagement estimation task by adding \textsc{PInSoRo}, English-language Child-Child and Child-Robot play sessions annotated with social and task engagement. Even though the additional subtask used classification instead of continuous annotation labels, several teams proposed new approaches or refined existing models that generalised knowledge across engagement label domains, improving upon last year's winning model and substantially outperforming the \textsc{PInSoRo} baselines. %

\begin{acks}
Marius Funk was supported by a scholarship of the German Academic Exchange Service (DAAD).
This work was supported in part by the Deutsche Forschungsgemeinschaft (DFG) through the Leibniz Award of Elisabeth André under Grant AN 559/10-1.
Michal Balazia was funded by the \grantsponsor{}{French National Research Agency}{https://anr.fr/} under the UCA\textsuperscript{JEDI} Investments into the Future, project number \grantnum{}{ANR-15-IDEX-01}.
This work has been partially supported by the German Federal Ministry of Research, Technology and Space (BMFTR) under the Robotics Institute Germany (RIG).
We thank S\'everin Lemaignan for his help with integrating the PInSoRo dataset in our challenge.

\end{acks}

\bibliographystyle{ACM-Reference-Format}
\balance
\bibliography{bibliography}

\end{document}